\documentclass[times, 10pt,twocolumn]{article}
\usepackage{latex8}

\usepackage{latexsym,color,graphics}
\usepackage{diagrams}
\usepackage{url}
\usepackage{epsfig}
\usepackage{amssymb}
\usepackage{amsmath}
\usepackage{amsfonts}

\begingroup
\catcode`\~=11
\gdef\urltilde{\lower 0.6ex\hbox{~}}
\endgroup

 \newcommand{\T}{\mathcal{T}}
 
\newcommand{\W}{\mathcal{W}}

\newtheorem{propo}{Proposition}

\newtheorem{definition}{Definition}

\begin{document}

\title{Temporal Probabilistic Logic Programs: State and Revision}


\author{Zoran Majki\'c \\
International Society for Research in Science and Technology,\\
 PO Box 2464 Tallahassee, FL 32316 - 2464 USA\\
majk.1234@yahoo.com, ~~~ http://zoranmajkic.webs.com/}
 \maketitle
\thispagestyle{empty}

\begin{abstract}
There are numerous applications where we have to deal with temporal
uncertainty associated with events.  The Temporal Probabilistic (TP)
Logic Programs  should provide support for \emph{valid-time
indeterminacy} of events, by proposing the concept of an
indeterminate instant, that is, an interval of time-points (event's
time-window) with an associated, lower and upper, probability
distribution. In particular, we propose the new semantics, for the
TP Logic Programs of Dekhtyar and Subrahmanian. Our semantics, based
on the possible world semantics is a generalization of the possible
world semantics for (non temporal) Probabilistic Logic Programming,
and we define the new syntax for PT-programs, with time variable
explicitly represented in all atoms, and show how the standard role
of Herbrand interpretations used as possible worlds for probability
distributions is coherently extended to Temporal Probabilistic Logic
Programming.
\end{abstract}

\section{Introduction}
The reasoning with probabilistic information based on PSAT
(Probabilistic Satisfiability)  is the problem of determining wether
a set of assignments of probabilities to a collection of boolean
formulas of atomic events is consistent has a long history
\cite{Bool54,Hail65,GeKP88}, and is proven that is NP-complete. But,
the probabilities derived from any sources may have tolerances
associated with them, and Fenstad \cite{Fens80} has shown  that when
enough information is not available about the interaction between
events, the probability of compound events
 cannot be determined precisely: one can only give bounds,
lower and upper probability bound. Consequently, the probability
intervals used for uncertain information are the simplest extension
of the traditional probability modes \cite{Wall91,CaHM94,Weic99} and
are used also as belief measure for uncertainty in fuzzy logics.
Such metric for non temporal logic programming (p-programs) is used
in a number
of papers \cite{NgSu92,LaSa94,LLRS97,DeDS99,DeSu00}.\\
 We assume that every event occurs at a point in time with a
 probability interval of reals $[a,b] \subseteq [0,1]$
 An instant (time point or chronon) $\textbf{t}$ is
specified  w.r.t a given time granularity of a linear calendar
structure $\T$; for example "$day/month/year$". Often, however, we
do not know the exact time point; instead, we only know that the
instant is located sometime during a time interval. We call such an
instant an \emph{indeterminate instant} ~\cite{DySn98}. Dyreson and
Snodgrass have drawn attention to the fact that, in many temporal
database applications, there is often uncertainty about the start
time of events, the end time of events, and the duration of events.
The indeterminacy refers to the \emph{time} when an event occurred,
not whether the event occurred or not. An indeterminate instant is
described by lower and upper time bound, and a probability
distribution (mass) function (PDF) ~\cite{DRSu01} which, for every
time point in this interval, returns with lower and upper
probability value assigned to a chronons. Generally, for the
interval-based lattice the first introduction of interval-based
Temporal Probabilistic Logic Programs (TP-programs) is presented in
\cite{DeDS99T}, and is extended to TP-databases \cite{DRSu01}, so
that the semantic of interval-based Probabilistic Logic Programs
based on possible worlds and the fixpoint semantics for such
programs \cite {NgSu92} is considered valid for more than 13 years.
But recently the author,
had the possibility to approach the general problems with such TP
databases \cite{UZVS05},  to consider the semantics of TP-Logic
Programs and to realize that it is not correctly defined.\\
The initial suspect for the validity of the
fixpoint semantics w.r.t the model theory of the p-programs
(Probabilistic programs), defined in the seminal paper \cite{NgSu92}
and successively repeated in all other papers, was based on the two
observations: on an unnatural semantics for the probability
interval-based bilattice used for computation of the fixpoint, and
on the intuition that would be possible to convert the p-programs
with interval-based annotated atoms into the probabilistic
constraint programs, and for them there is no guarantee that the
solution will contain only simple probabilistic intervals for atoms.\\
The first consideration was analyzed and presented  in \cite{Majk05Fu}; briefly:\\
In the bilattice $L_B$ of (closed) probability intervals  we
associate to each fact of knowledge database the belief measure
$[x,y] \in L_B$. Such belief is \emph{consistent} if $x\leq y$, that
is, when the lower boundary is less than upper boundary.\\
The \emph{belief} (or truth) ordering in $L_B$ is defined as
follows: $~~[x,y] \leq_B [x_1,y_1]$ $~~~$ iff $~~~$ $x \leq x_1$ and
$y \leq y_1$. It means that the belief $[x_1,y_1]$ is higher than
the belief $[x,y]$, that is,  for any probability $a$ which
satisfies the first belief (i.e., $a \in [x_1,y_1]$) there is a
probability $b$ that satisfies the second belief such that $a \leq
b$. The element $0_B = [0,0]$ and $1_B = [1,1]$ are the bottom and
the top element of
this ordering.\\
There exists also the \emph{precision} (or knowledge) ordering in $L_B$ and is defined as follows: \\
$~~~$$[x,y] \leq_K [x_1,y_1]$ $~~~$ iff $~~~$ $x \leq x_1$ and $y
\geq y_1$. It means that the belief $[x_1,y_1]$ is more precise than
the belief $[x,y]$. The join and meet operations for this ordering,
$\wedge_{kn}, \vee_{kn}$, respectively,
 are defined as follows:
 $~~~$$~[x,y] \wedge_{kn} [x_1,y_1] = [min\{x,x_1\}, max\{y,y_1\}
 ]$,
 $~~~$$~[x,y] \vee_{kn} [x_1,y_1] = [max\{x,x_1\}, min\{y,y_1\} ]$.
 The probabilistic interpretation of conjunction and disjunction
 correspond to the ignorance strategy, that is
 $~[x,y] \wedge_{ig} [x_1,y_1] =
[max\{0, x+x_1 -1\}, min\{y,y_1\} ]$, $~[x,y] \vee_{ig} [x_1,y_1] =
[max\{ x,x_1\},
min\{1,y+y_1\} ]$,\\
which are not meet/join lattice operators for any of these two
orderings (usually the fixpoint computation uses the meet/join
lattice operators). In \cite{NgSu92} is used the knowledge
(precision) ordering for a computation of the least
 fixpoint semantics for interval-based  logic
 programs, but the disjunction $\vee_{kn}$ can produce as result inconsistent values $[x,y]$ such that $x>y$; moreover, while to the
 bottom solution is reasonably assigned the whole interval $0_K = [0,1]$ (bottom value of the lattice), to the 'best solution' is assigned the top value
 of the lattice $1_K = [0,1]$ (denominated empty interval $\emptyset$ in \cite{NgSu92}) which is \emph{inconsistent}
  (i.e., the best solutions result inconsistent)!\\
  Also the second observation was investigated by the author and the result
  is presented in
\cite{Majk05E} with the reduction of TP-databases into Constraint
Logic Programs: consequently, we are able to apply interval PSAT in
order
to find the models of such interval-based probabilistic programs, but,  such models cannot,
in general case, be described by single intervals associated with atoms of a program.\\
In fact, the author discovered that there are cases when the least
 fixpoint of a p-program P \emph{is not} model of P: it happens
 always when there is a rule with an atom in the body with the probability
 interval more thin then the interval for this atom assigned by the
 least fixpoint (so that this rule could not be satisfied during the
 least fixpoint computation) and with the atom in the head of this
 rule with the probability interval more thin that the interval for
 this atom computed by the least fixpoint.
 In order to make serious revision for TP-programs and their
 semantics, we needed some additional mathematical tools also, based on
 the concepts of the predicate compression and Higher-order
 Herbrand model types \cite{Majk06} (used also
 for 'abstracted' databases in
 \cite{Majk05E}).\\
\textbf{Remark}: Independently form this author's investigation
about the validity of the given semantics for interval-based
p-programs, and presumably in the same time  (such coincidence is
astounding), also the two coauthors which previously worked on this
issue for more than 5 years \cite{DeDS99,DeDS99T}, discovered the
incorrectness of their previous definition of fixed point semantics
for interval-based p-programs. So that in the first paper
\cite{DeDe04} they presented the contra examples, and proved that
the fixpoint semantics of p-programs is \emph{unsound} (their
Proposition 1 shows that the fixpoint semantics derives the
interpretations which are not models of a p-program) and
\emph{incomplete} (their Proposition 3 shows that there are models
of a p-program which are not interval-based, so that the fixpoint
operator is unable to find any solution). The correct semantics for
probabilistic programs (p-programs), based on interval PSAT, is
presented recently \cite{DeDe05} and shows that the entailment
problem for p-programs is co-NP-complete. Only for the particular
subset of p-programs (denominated \emph{simple strict} programs in
\cite{DeDe05}) this complexity is
polynomial and can be computed by the original fixpoint semantics defined in \cite{NgSu92}.\\
But in this brief history for the relevant work, the TP-programs,
which are more complex than p-programs, and  which have taken the
principal attention of the author during his collaboration for the
definition of algebra for TP-databases with aggregations, where not
taken into right
consideration. This is the main aim of this paper. \\
The author's opinion is that the main drawbacks of the work in
\cite{DeDS99T} can be summarized as follows:  its fixpoint semantics
is incorrect w.r.t. its model theory in the analog way described
above in the case of more simple p-programs; the second is based on
the fact that its semantics for probability, based on possible
worlds, is apparently taken without any plausible connection with
the \emph{standard semantics} for p-programs based on Herbrand
interpretations, principally because they did not explored the
possible reductions of TP-programs into p-programs (similarly as in
the case when in \cite{NgSu92} was not considered the possible
reduction into Constraint Logic Programs, and the price was the
incorrect fixpoint semantics). Based on these observations, the more
important contributions in this work w.r.t. the work in
\cite{DeDS99T} can be summarized as
follows:\\
1. The definition of the new syntax  and the model theory for
Temporal Probabilistic programs, denominated PT-programs
(Probabilistic Temporal Programs), where is considered the full
temporal property of events by including the attribute for
time-points inside of all atoms (basic events): such atoms will be
denominated t-atoms in what follows. By this intuitive and simple
operation we obtain t-Herbrand models and indirectly the reduction
of PT-programs into the p-programs with t-atoms, so that the
possible world semantics for the PT-programs with this new syntax is
based on  standard Herbrand models.\\
2. Such new PT-programs has the same possible world semantics for
p-programs \cite{DeDe05} which can be solved, in the general case,
by interval PSAT as discussed in precedence.\\
3. We show how these PT-programs can be transformed in the previous
version of TP-programs described in \cite{DeDS99T}, by means of
predicate compression for the temporal attribute: thus, the possible
worlds of old TP-programs is is the set of Higher-order Herbrand
interpretations which are result of this predicate compression. The
TP-programs obtained by this transformation (which is knowledge
invariant)  do not suffer the  semantics drawback as in
\cite{DeDS99T}, and can be considered as the minimal revision of the
work presented in \cite{DeDS99T}.\\
The plan of this work is the following: After brief introduction in
invariant flattening-compression knowledge transformation, in
Section 3 we introduce the new syntax for temporal probabilistic
logic programs (PT-programs) with t-atoms and more expressive
interval-based probabilistic annotation w.r.t. the definition in
\cite{DeDS99T}. In Section 4 we develop the model theoretic
semantics for PT-programs, by reduction to ordinary probabilistic
p-programs, and we define the complexity for consistency and
entailment problem for PT-programs. In Section 5 we make comparison
of the new PT-program's semantics and the model theoretic semantics
for TP-programs given in \cite{DeDS99T}: we show their coincidence,
and explain that the possible worlds in \cite{DeDS99T} are the
higher-order Herbrand interpretation obtained by compression of
temporal variable in PT-programs.\\ Finally, in Section 6 we apply
the PT-programming for the evolution in time, which modify only
p-annotations, of ordinary (non temporal) p-programs.  We also
discuss the future work and the challenges for effective
query-answering in PT-programming for Temporal Probabilistic
Databases. $\vspace*{-2mm}$
%
\section{Invariant Knowledge transformation: Flattening and
compression duality}  \label{sec:Krip}
The higher-order Herbrand interpretations of logic programs (for
example Databases) \cite{Majk06}, produce models where the true
values for ground atoms are not truth constants but functions. In
this section we will give the general definitions for such
higher-order Herbrand interpretation types for logic programs and
their models. More detailed information can be found in
\cite{Majk06FM}.\\ We denote by $A\Rightarrow B$, or $B^A$, the set
of all functions from $A$ to $B$, and by $\textbf{2} = \{0,1\}$ the
set of logic values (0 for the false, 1 for the true logic value).
\begin{definition} (Higher-order Herbrand interpretation types \cite{Majk06}): $~~~ $ Let $H^{com}$ be a Herbrand
base, then, the higher-order Herbrand interpretations are defined by
$~I_{com}:H^{com}\rightarrow T~$, where $T $ denotes the functional
space $W_1\Rightarrow (...(W_n \Rightarrow \textbf{2})...)$, denoted
also as $(...((\textbf{2}^{W_n})^{W_{n-1}})...)^{W_1}$, and $W_i, ~
i \in [1,n],~n\geq 1 $, the sets of parameters.\\
In the case  $n = 1$,  $T = (W_1\Rightarrow \textbf{2})$, we will
denote this interpretation by $~I_{com}:H^{com}\rightarrow
\textbf{2}^{W_1}~$.
\end{definition}
The  interpretations $I_{com}:H^{com}\rightarrow \textbf{2}^{\W}$
are higher-order types of Herbrand interpretations: the set of truth
values for them are \emph{functions} instead of constants. We pass
from a flat truth structure for atoms in a Hebrand interpretations
 to non flat functional space truth
structure for atoms in the compressed Herbrand base $H^{com}$.
\\ Now we will
introduce the top-down  transformation, called \emph{flattening},
where the context (uncertain or approximated information), defined
as the set  $\W$, is fused into the Herbrand base by enlarging
original predicates of old theory with new attributes taken from the
context. In this way the hidden information of the context becomes a
visible information and a visible part of the logic language. In
what follows, for any given k-ary predicate symbol $r$, for a given
tuple of constants $\textbf{d} = <d_1,..,d_k>$ in a given Herbrand
universe,   $r(\textbf{d})$ denotes a ground atom of the resulting
Herbrand base $H^{com}$.
\begin{definition}  (Flattening - Global
decompression \cite{Majk06FM})\label{def:flat}\\
Each higher-order Herbrand interpretation
$~I_{com}:H^{com}\rightarrow T~$, where $T $ denotes the functional
space $W_1\Rightarrow (...(W_n \Rightarrow \textbf{2})...)$, and
$~\W = W_1\times...\times W_n$ cartesian product, can be flattened
into the  Herbrand interpretation $~~~I_F:H_F\rightarrow
\textbf{2}$, $~~~$ where $~~~~H_F =
\{r_F(\textbf{d},\textbf{w})~|~r(\textbf{d}) \in H^{com} $ and
$\textbf{w} \in \W \}$,\\
 is the Herbrand base of new predicates $r_F$, obtained as extension
 of original predicates $~r~$ by parameters (attributes
 for domains in $W_i, 1 \leq i \leq n$),
such that for any $~~r_F(\textbf{d},\textbf{w})\in H_F,~\textbf{w} =
(w_1,...,w_n) \in \W$, holds that
$~~~~I_F(r_F(\textbf{d},\textbf{w})) =
 I_{com}(r(\textbf{d}))(w_1)...(w_n)$.
\end{definition}
 We define as \emph{parameterizable} Herbrand base any
 Herbrand base such that \emph{all} its atoms have the \emph{common set}
 of attributes $\textbf{y} = \{y_1, y_2,..,y_k\}$. In this, most
 simple case of compression, we can obtain a compressed  Herbrand base,
 denoted by $H^{com}$, in the way that these common
 attributes become hidden attributes.
 \begin{definition} (Global compression \cite{Majk06FM})\label{def:abs}.\\ Let
 $I_F:H_F\rightarrow \textbf{2}$ be the 2-valued Herbrand interpretation for
a parameterizable  Herbrand base $H_F$.
 Then the interpretation for its compressed Herbrand base\\
$~~H^{com} = \{r(\textbf{d})~|~ r_F(\textbf{d},\textbf{w}) \in
H_F\}$ is defined by $~~~~~~I_{com}:H^{com}\rightarrow
\textbf{2}^{\W}$, $~~~~~~$ such that $I_{com} = [I_F\circ is]$,
 where $\W = Dom_{y_1}\times...\times Dom_{y_k}$ is the set of all
 parameter tuples.
This bijective correspondence of $I_F$ and $I_{com}$ is given by the
following commutative diagram
\begin{diagram}
   \textbf{2}^{\W}\times \W  & \rTo^{eval}   & \textbf{2} \\
 \uTo^{I_{com} = [I_F \circ is]} \uTo_{id_W}& \ruTo_{I_F\circ is} & \uTo_{I_F} \\
 H^{com} \times \W  &  \rTo^{is}    & H_F \\
\end{diagram}
where $is:H^{com}\times \W \simeq H_F$ is a bijection, such that for
any $r(\textbf{d})\in H^{com}, w \in \W$, $is(r(\textbf{d}),w) =
r_F(\textbf{d},w)$, $[\_]$ is the curring ($\lambda$ abstraction)
for functions, $\textbf{2}^{\W}$ is the set of functions from $\W$
to $\textbf{2} = \{0,1\}$, and $eval$ is the evaluation of the
function in $\textbf{2}^{\W}$ for the values in $\W$.
\end{definition}
No one of subsets $S\subseteq H^{com}$ can be a \emph{model} for a
compressed database; that is, the models for compressed database
\emph{are not} ordinary Herbrand models but a kind of
\emph{higher-order type}
of Herbrand models.\\
%
%
%
%
\section{New Probabilistic Temporal programs: Syntax of PT-programs}
We will use the same terminology as in \cite{DeDS99T}. The main
difference is that our event atoms, differently form event atoms
\cite{DeDS99T} have also the temporal attribute $y$ with a domain
represented by  $S_\tau$, the set of all valid time points
$\textbf{t} \in S_\tau$ of a calendar of a type $\T$. For example,
let $p(\textbf{d})$ be an ordinary atom with an n-ary predicate
symbol $p$ and a tuple of n constants or variables $\textbf{d}$.
Then $A = r_F(\textbf{d},y)$, where $r_F$ is n+1-ary predicate
symbol obtained from $p$ by enlarging it with a new temporal
attribute, is an event's t-atom (temporal-atom). When the tuple
$\textbf{d}$ is composed by only constants and $y$ is a time point
in $S_\tau$ then $A$ is said to be ground t-atom. If $A_1,..,A_k$
are the (simple)  t-atoms, then $A_1 \wedge...\wedge A_k$ and $A_1
\vee...\vee A_k$
are called compound  t-atoms.\\
 Let $L$ be a language generated for compound events by a given set of constants (Herbrand universe) and temporal predicate
 symbols.
We  assume that all variable symbols from $L$ are partitioned into
three classes: the \emph{object variables} (contains the regular
first order logic variable symbols: variables in a tuple
$\textbf{d}$ of the example above), the \emph{probabilistic}
variables (range over the interval of reals in $[0,1]$) and
\emph{temporal} variables (range over the set of time points
$S_\tau$ of a given calendar: in the examples we will use integer
numbers for time points): the temporal variable $y$ in t-atoms will
be called \verb"principal" variable $y$ and all other temporal
variables will be called independent.\\
Let $Var$ be a set of variables and $S$ be a set of constants. The
terms are defined as follows:\\
1. all variables $x \in Var$, and constants $d \in S$ are terms;\\
2. if $f:S^n \rightarrow S$ is a functional symbol of arity $n$ and
$\lambda_1,..,
\lambda_n$ are terms, then $~f(\lambda_1,..,\lambda_n)$ is a term.\\
We define two types of terms: the temporal terms, when $S = S_\tau$;
and probabilistic terms, when
 $S = [0,1]$.\\
\textbf{Temporal Constraint}: A temporal constraint
$C = c(y,y_1,...,y_k)$ with principal variable $y$ and other variables $y_1,...,y_k$ is defined inductively: \\
1. let $\lambda$ be a temporal term with the set of variables
$y_1,...,y_k$, then $~~(y~ op~\lambda)$, where $op \in \{
\leq,<,=,\neq,>,\geq \}$, is a temporal constraint. The $y:\lambda_1
\thicksim \lambda_1$ is a short denotation for
 $y \geq \lambda_1 \wedge y \leq \lambda_1$.\\
2. if $C_1$ and $C_2$ are temporal constraints with the same
principal variable $y$, then $C_1\wedge C_2,~C_1\vee C_2$, and $\neg
C_1$ are temporal constraints.\\
 A temporal constraint is called normal if it does not contain
 variables different from the principal variable.\\
Let $C = c(y)$ be a normal temporal constraint, the the
\emph{solution set} of time points of $C$ is equal to $sol(C) =
\{\textbf{t}~|~\textbf{t} \in S_\tau$ and $c(\textbf{t})$ is true
$\}$, with the cardinality $|sol(C)|$.\\
\textbf{Probabilistic weight function}: for any given temporal
constraint $C = c(y,y_1,...,y_k)$, we define the function\\
$\omega_C:S_\tau^{k+1} \rightarrow [0,1]$, such that for any
$\{\textbf{t}, \textbf{t}_1,...,\textbf{t}_k\} \in S_\tau^{k+1}$, if
$~\omega_C(\textbf{t}, \textbf{t}_1,...,\textbf{t}_k) \neq 0~$ then
$~\textbf{t} \in sol(C)$.\\
Alternatively, in the case when $|sol(C)| = m$ is a finite number,
we will specify the weight function in the form of the time-ordered
set of values $\{v_1,...,v_m\}$: For example, if $sol(c(y)) =
\{\textbf{t}_1,..,\textbf{t}_3\} $, a weight function $\omega_C$ can
be represented as $\{0.4, 1, 0.5\}$, and it will mean that
$\omega_C(\textbf{t}_1) = 0.4, \omega_C(\textbf{t}_2) = 1,
\omega_C(\textbf{t}_3) = 0.5$. We will denote by $\sharp$ the  constant weight function (equal to 1) for the constraints with $|sol(C)| = 1$.\\
 The intuition underlying
the above definition is that a probabilistic weight function
$\omega_C$, of a given temporal constraint $C$, assigns a
probability $p \geq 0$ to each time point in the solution set of
this temporal constraint (for all other time
points it must be equal to zero).\\
\textbf{Temporal probabilistic annotation}: a tp-annotation  is a
triple $\langle C,  \omega_{C_L}, \omega_{C_U} \rangle$ where $C$ is
temporal constraint, $\omega_{C_L}$ and $\omega_{C_U}$ are
probabilistic weight functions for lower and upper boundary respectively.\\
\textbf{Remark}: this is more general definition then in
\cite{DeDS99T}, but gives us possibility to model lower and upper
probability boundaries independently.
 \begin{definition} Let $F = A_1 \ast ...\ast A_k$ be a  compound event
 t-atom, where $\ast \in \{\wedge, \vee\}$, and $\mu = \langle C,  \omega_{C_L}, \omega_{C_U} \rangle$
  be a tp-annotation, then $~F:\mu~$ is a tp-annotated basic
 formula.\\
 Let $A:\mu, F_1:\mu_1,..., F_m:\mu_m$ be tp-annotated basic
 formulae and $A$ a t-atom. Then $~A:\mu \leftarrow F_1:\mu_1 \wedge...\wedge
 F_m:\mu_m$ is a tp-clause.\\
 A \verb"Probabilistic Temporal Program" (PT-program) is a finite set of
 tp-clauses. If $P$ is a PT-program, we let $ground(P)$ denote the
 set of all ground instances of rules of P. By $H_F$ we denote the
 Herbrand base of a program P for a given set of  constants for
 object variables.
  \end{definition}
  \textbf{Remark}: as we can see this syntax is simile to the syntax for
  TP-programs presented in \cite{DeDS99T}. The main difference is
  that all atoms in our definition of PT-programs are
  \emph{t-atoms}: as the consequence we will have that the temporal
  constraint in tp-basic formulae is an 'internal' annotation for
  the t-atoms (the temporal attribute of any t-atom corresponds to the dependent variable of the temporal constraint),
   while the probabilistic annotation remains an external (standard)
  annotation for t-atoms.\\
  To underlay these simple modification we will use the same example
  (Example 2 presented in \cite{DeDS99T}), but with a new syntax for
  PT-programs:\\
 \textbf{ Example 1}: For a company which deals with projected
 arrivals of the packages shipped by the company. First two rules
 provide the information on the probability distribution of the
 arrival time of an arbitrary package sent to any place. The third
 rule gives some extra information about the arrival time of
 packages sent to Paris via express-mail. Three facts about
 shipments complete this program.\\
 $arrived_F(Item,Place,y):\langle y:3 \thicksim 5,
 \{.25,.15,.1\},\\ ~~~~~~~~~~~~~~~~~~~~~~~~~~~~~~~~~~~~~~~~~~~~~~~~~~~~~~\{.4,.24,.16\}\rangle \\
 ~~~~~~~~~ \leftarrow   sent_F(Item,Place,y):\langle y=1,
  \{0.9\},\sharp\rangle$,\\
$arrived_F(Item,Place,y):\langle y:6 \thicksim 8,
\{.15,0,.05\},\\~~~~~~~~~~~~~~~~~~~~~~~~~~~~~~~~~~~~~~~~~~~~~~~~~~~~~~\{.3,0,.1\} \rangle \\
  ~~~~~~~~~\leftarrow   sent_F(Item,Place,y):\langle y=1,
  \{0.9\},\sharp\rangle$,\\
  $arrived_F(Item,Place,y):\langle y:3 \thicksim 4,
  \{.3,.2\},\\~~~~~~~~~~~~~~~~~~~~~~~~~~~~~~~~~~~~~~~~~~~~~~~~~~~~~~  \{.54,.36\} \rangle \\
  ~~~~~~~~~\leftarrow  sent_F(Item,\textsl{paris},y):\langle y=1,
  \{.95\},\sharp\rangle  \\ ~~~~~~~~~~~~\wedge  express-mail_F(Item,y):\langle y=1,
  \sharp,\sharp\rangle$ ,\\
$sent_F(\textsl{shoes},\textsl{rome},y):\langle y=1,
  \sharp,\sharp\rangle \leftarrow $,\\
  $sent_F(\textsl{letter},\textsl{paris},y):\langle y=1,
  \sharp,\sharp\rangle \leftarrow $,\\
  $express-mail_F(\textsl{letter},y):\langle y=1,
  \sharp,\sharp\rangle \leftarrow $.

%
\section{Model Theory for PT-programs} \label{sec:PT}
In this section we will show that each PT-program $P$ has the
standard probabilistic model theory based on the Herbrand base $H_F$
of $P$, with the set of possible worlds equal to the set $I_F \in
\textbf{2}^{H_F}$ of Herbrand interpretations of $P$, $I_F:H_F
\rightarrow \textbf{2}$, where $\textbf{2} = \{0,1\}$ is the set of
logic values (0 for the false, 1 for the true logic value). Each
model theory assumes that in real world each t-atom in $H_F$ is
either true or false. In our case, in any possible world $I_F \in
\textbf{2}^{H_F}$, for any t-predicate symbol $r_F$ and the tuple of
constants $\textbf{d}$ of its object variables, the set of time
points $\{\textbf{t}_i~|~r_F(\textbf{d},\textbf{t}_i) \in H_F$ and
$I_F(r_F(\textbf{d},\textbf{t}_i)) =1\}$ corresponds to the temporal
uncertainty of this event: in each time point of this set the
event's uncertainty is bounded in a form of a probability
interval.\\
A variable assignment $\sigma$ maps each object variable to an
object constant and each temporal variable to the set $S_{\tau}$ of
time points of the calendar. The truth of the events $\phi \in L$ in
$I_F$ under $\sigma$, denoted by $I_F \models_{\sigma} \phi$, is
inductively defined as follows:\\
1. $I_F \models_{\sigma} r_F(a_1,..,a_k,y)~~$ iff
$~~I_F(r_F(\sigma(a_1),.., \sigma(y))) = 1$, $~~~~$ for every t-atom $r_F(a_1,..,a_k,y)$;\\
2. $I_F \models_{\sigma} \phi \wedge \psi ~~$ iff $~~I_F
\models_{\sigma} \phi$ and $I_F \models_{\sigma} \psi$;\\
3. $I_F \models_{\sigma} \phi \vee \psi ~~$ iff $~~I_F
\models_{\sigma} \phi$ or $I_F \models_{\sigma} \psi$.\\
An event $\phi$ is true in a possible world $I_F$, or $I_F$ is a
model of $\phi$, denoted $I_F \models \phi$, iff $I_F
\models_{\sigma} \phi $ for all variable assignments $\sigma$.\\
 In order to be able to apply the results of the
standard possible world semantics for PT-programs, we have to
 show that each
PT-program corresponds to \emph{standard probabilistic} program
(p-program).
\begin{propo} Each PT-program is a pure Probabilistic Logic   Program.
\end{propo}
\textbf{Proof}: It can be shown by simply unfolding of the temporal
constraints in tp-annotated basic formulae, that is, by partial
grounding of the temporal attributes of t-atoms in a given
PT-program. That is, given a tp-clause $~A:\mu_0 \leftarrow
\bigwedge_{1 \leq k \leq m } F_k:\mu_k $, with the t-atom $A =
r_F(\textbf{v},y)$ with a tuple of object variables e/o constants in
$\textbf{v}$ and $\mu_k = \langle
C_k,\omega_{C_{L_k}},\omega_{C_{U_k}}
 \rangle, 0 \leq k \leq m$, we can unfold this tp-clause in the
following finite set (because the calendar is finite) of
p-clauses:\\
$\{ r_F(\textbf{v},\textbf{t}_i):[ \omega_{C_{L_0}}(\textbf{t}_i),
 \omega_{C_{U_0}}(\textbf{t}_i) ]~~ \leftarrow \\
~~~~~~~~~~~~~~\bigwedge_{1 \leq k \leq m }(\bigwedge_{\textbf{t}_j
\in
sol(C_k)} \Phi_k(\textbf{t}_j))~|~\textbf{t}_i \in sol(C_0) \}$,\\
where $~\Phi_k(\textbf{t}_j) = F_k(\textbf{t}_j):[
\omega_{C_{L_k}}(\textbf{t}_j),  \omega_{C_{U_k}}(\textbf{t}_j) ]$,
and $F_k(\textbf{t}_j)$ is obtained from the $F_k$ by substitution
of the temporal variable in t-atoms of $F_k$ by the
constant (time point) $\textbf{t}_j$.\\
Thus, we obtain a p-program where all annotations of basic
p-formulae are constant probabilistic intervals.\\
$\square$\\
\textbf{ Example 2}: Let us consider the PT-programs of the Example
1. The first tp-clause of this programs will be unfolded into the
set
of the following three p-rules:\\
$arrived_F(Item,Place,3):[.25, .4] \\
 ~~~~~~~~~ \leftarrow   sent_F(Item,Place,1):[.9, 1]$,\\
 $arrived_F(Item,Place,4):[.15, .24] \\
 ~~~~~~~~~ \leftarrow   sent_F(Item,Place,1):[.9, 1]$,\\
 $arrived_F(Item,Place,5):[.1, .16] \\
 ~~~~~~~~~ \leftarrow   sent_F(Item,Place,1):[.9, 1]$,\\
 Notice that such simple transformation for the definition of the
 TP-programs in \cite{DeDS99T} is impossible because their version
 does not include the temporal attribute in event's atoms.\\
$\square$\\
 Thus, given Herbrand base $H_F$ of a PT-program $P$
(equal to the Herbrand base of the p-program obtained by the
unfolding described above), a world probability density function
$KI$ is defined as $KI:\textbf{2}^{H_F} \rightarrow [0,1]$, such
that for all $I_F \in \textbf{2}^{H_F}, ~KI(I_F) \geq 0$ and
$\sum_{I_F \in \textbf{2}^{H_F}} KI(I_F) =
1$ (Kolmogorov axioms).\\
A \emph{probabilistic interpretation} (p-interpretation)\\ $~I:H_F
\rightarrow [0,1]$ of a PT-program $P$ is defined  as follows:\\
$I(A) = \sum_{I_F \in \textbf{2}^{H_F}} I_F(A)\cdot KI(I_F)$, $~~$ for any ground t-atom $A \in H_F$.\\
That is, p-interpretation assigns probabilities to individual ground
t-atoms of $H_F$ by adding up the probabilities of all worlds $I_F$
in which a given t-atom is true (i.e., $I_F(A) = 1$).\\
Given a p-interpretation $I$, it can be extended to all compound
events in $L$ by the mapping $Pr:L\rightarrow [0,1]$, such that the
probability of an event $\phi$ in the probabilistic interpretation
$Pr$ under a variable assignment $\sigma$, denoted
$Pr_{\sigma}(\phi)$, is the sum of all $KI(I_F)$ such that $I_F \in
\textbf{2}^{H_F}$ and $I_F \models_{\sigma} \phi$ (we write
$Pr(\phi)$ when $\phi$ is
ground), that is\\
$~~~~Pr_{\sigma}(\phi) = \sum_{I_F \in ~\textbf{2}^{H_F}, ~I_F
\models_{\sigma} \phi} ~KI(I_F)$.\\
p-interpretations specify the model-theoretic semantics of
p-programs, as follows:\\\\
1. $Pr \models_{\sigma} F:[a,b]~~$ iff $~~Pr_{\sigma}(F) \in [a,b]$,
$~~~~~~$ that is, $~~~~~~~~~~ $ iff $~~ a \leq Pr_{\sigma}(F) \leq b$;\\
 2. $Pr \models_{\sigma} F_1:[a_1,b_1] \wedge ... \wedge F_n:[a_n,b_n]~~ $ iff $~~(\forall 1 \leq i \leq n)(Pr
\models_{\sigma} F_i:[a_i,b_i])$;\\
3. $Pr \models_{\sigma} F:[a,b] \leftarrow F_1:[a_1,b_1] \wedge ...
\wedge F_n:[a_n,b_n]~~$ iff $~~Pr \models_{\sigma} F:[a,b]$ or
$Pr \nvDash_{\sigma} F_1:[a_1,b_1] \wedge ... \wedge
F_n:[a_n,b_n]$.\\
As we can see, from the point 1 above, the satisfaction of
p-programs is based on the Interval PSAT for the system of
inequalities: any assignment by $I$ (that is, $Pr$) of point
probabilities to the atoms, that satisfies these \emph{constraints}
is a model of $P$.\\
Now we are ready to specify the model-theoretic semantics for
PT-programs, as follows:
\begin{definition}(\textbf{Satisfaction}) \label{def:satif} Let $\sigma$ be an assignment only for object variables, then\\
1. $I \models_{\sigma} F:\mu~~~~$ iff $~~(\forall \textbf{t} \in
sol(C))( Pr
 \models_{\sigma} F(\textbf{t}):[\omega_{C_L}(\textbf{t}),
\omega_{C_U}(\textbf{t})])$,\\
where $\mu = \langle C,\omega_{C_L},\omega_{C_U}\rangle$ and
$F(\textbf{t}))$ is obtained from $F$ by substitution of the
temporal variable in
t-atoms of $F$ by the constant (time point) $\textbf{t}$;\\
2. $I \models_{\sigma} F_1:\mu_1 \wedge ... \wedge F_n:\mu_n~~~~ $
iff $~~(\forall 1 \leq i \leq n)(I
\models_{\sigma} F_i:\mu_i)$;\\
3. $I \models_{\sigma} A:\mu \leftarrow F_1:\mu_1 \wedge ... \wedge
F_n:\mu_n~~~~$ iff $~~I \models_{\sigma} A:\mu~$ or $~I
\nvDash_{\sigma} F_1:\mu_1 \wedge ... \wedge F_n:\mu_n$.
\end{definition}
A tp-clause $Cl$ is true in a probabilistic interpretation $I$ (that
is, in its extension $Pr$), or $I$ is a model of $Cl$, denoted $I
\models Cl$, iff $Pr\models_{\sigma} Cl$ for all object variable
assignments $\sigma$. \\$I$ is a model of a PT-program $P$ if it is
a model for all tp-clauses in $P$. Let $Mod(P)$ denote the set of
all models of a PT-program $P$; $P$ is called \emph{consistent} iff
$Mod(P) \neq \emptyset$, otherwise $P$ is called
\emph{inconsistent}.\\ A PT-program $P$ is \emph{satisfiable} iff a
model of $P$ exists.\\ A tp-annotated basic formula $F:\mu$ is a
\emph{logical consequence} of a PT-program $P$, or $P$ entails
$F:\mu$, denoted $P \models F:\mu$, iff each model of $P$ is also
model of $F:\mu$.
\begin{propo}  The consistency problem for PT-programs is NP-complete, while the entailment problem for PT-programs is
co-NP-complete.
\end{propo}
It derives from the reduction of PT-programs into ordinary
p-programs, and form the complexity of interval PSAT for linear
inequalities (see Th.4.11 in\cite{FaHM90} and Th.3 in
\cite{DeDe05}).\\
By this results we have shown that the fixpoint semantics for
TP-programs, as defined in \cite{DeDS99T} is incorrect, that is
unsound and incomplete, similarly to the simpler case of the
fixpoint semantics of p-programs defined in \cite{NgSu92} and
propagated in
the dozen of the papers published after this seminal paper.\\
Instead, in what follows we will try to save the part of the work in
\cite{DeDS99T} which is correct and can be alternatively used for
the temporal probabilistic programming.
\section{Comparison of TP and PT Model theories}
As we can easy verify, the definition of the satisfaction relation,
for PT-programs given in  Definition \ref{def:satif} and for
TP-programs given in \cite{DeDS99T}, syntactically is equivalent
(consider that the point 1 of the Definition  \ref{def:satif} can be
reduced to\\
$I \models_{\sigma} F:\mu~~~~$ iff $~~(\forall \textbf{t} \in
sol(C))( Pr_{\sigma}(F(\textbf{t}))\in [\omega_{C_L}(\textbf{t}),
\omega_{C_U}(\textbf{t})])$,\\
which is syntactically equivalent to the point 1 of the definition
in \cite{DeDS99T}. But the $Pr$ used in Definition \ref{def:satif}
is based on the probabilistic interpretation $I:H_F \rightarrow
[0,1]$ where the set of possible worlds is the set of Herbrand
interpretations (as in standard world-based probability model
theory), while in \cite{DeDS99T} on the \emph{thread} function
$th:B_L \rightarrow \textbf{2}^{S_{\tau}}$, with the set of possible
worlds equal to the set of all threads (without the clear
explanation what is the connection with the standard probabilistic
model theory).\\
In what follows we will show that also their model theory is
correct, w.r.t. the different syntax for TP-programs, and we will
show how the set  of threats used for possible worlds is canonically
derived from the standard model theory of PT-programs (reducible to
pure p-programs) defined in Section \ref{sec:PT}.\\
Let $P$ be a PT-program with t-atoms, the Herbrand base $H_F$ and
the Herbrand model $I_F:H_F \rightarrow 2$. Then by the \emph{global
compression}, described in Definition \ref{def:abs}, for the
temporal attribute of all t-atoms in $P$, we obtain the TP-program
$P^{comp}$ with compressed atoms which contain only object
variables, and with the higher-order Herbrand model\\
$~~I_{com}:H^{com}\rightarrow \textbf{2}^{\W}$, $~~~$ such that
$I_{com} = [I_F\circ is]$,  $\W = S_{\tau}$, \\with the Herbrand
base $H^{com} = \{r(\textbf{d})~|~\exists \textbf{w}.
r_F(\textbf{d},\textbf{w}) \in H_F\}$, that is, $H^{com} = B_L$ and
$I_{com} = th$ . The diagram in Definition \ref{def:abs} is as
follows:
\begin{diagram}
   \textbf{2}^{S_{\tau}}\times S_{\tau}  & \rTo^{eval}   & \textbf{2} \\
 \uTo^{th = [I_F \circ is]} \uTo_{id_W}& \ruTo_{I_F\circ is} & \uTo_{I_F} \\
 B_L \times S_{\tau}  &  \rTo^{is}    & H_F    \\
\end{diagram}
Thus the set of threads $TH$ for the obtained TP-program $P^{comp}$
corresponds to the set of higher-order Herbrand interpretations,
i.e.,  $TH = (\textbf{2}^{S_{\tau}})^{B_L}$, bijective with the set
of possible worlds of the PT-program $P$, that is,
$(\textbf{2}^{S_{\tau}})^{B_L}~ \simeq ~\textbf{2}^{H_F}$, so that
the probability density function $KI$ for $P$ and $P^{comp}$ is the
same, and the satisfaction relation for $P^{comp}$ is identical to
the satisfaction relation for $P$.\\
Consequently the model-theoretic semantics for TP-programs, defined
in \cite{DeDS99T}, is equivalent to the model-theoretic semantics
for PT-programs defined in this paper. That is, given a ground atom
(with only object variables) $A =r(\textbf{d})$ in $B_L$, the
probability of this atom $p_M(r(\textbf{d}),\textbf{t})$ in a given
point of time $\textbf{t}$ \emph{is equal} to the probability of the
ground atom $r_F(\textbf{d},\textbf{t}) \in H_F$, as we can verify\\
$p_M(r(\textbf{d}),\textbf{t}) = \sum_{th \in TH,
  th(r(\textbf{d})) \ni \textbf{t}}KI(th)~$ (from Def. in \cite{DeDS99T})\\
  $= \sum_{I_F \in \textbf{2}^{H_F},~I_F(r_F(\textbf{d},\textbf{t}) =1}
  KI(I_F)~~~$ (from bijection $th \equiv I_F)$\\
  $ =  \sum_{I_F \in \textbf{2}^{H_F}}~I_F(r_F(\textbf{d},\textbf{t})\cdot KI(I_F) \\ =
  I(r_F(\textbf{d},\textbf{t}))$.\\

  So, from this point of view, both syntactical versions for
  temporal probabilistic logic programming can be used for
  applications: to have or not visible the time variable of events
  directly in the atoms of logic programs is left to the user's
  choice. With the semantic revision of the old syntax version in
  \cite{DeDS99T}, and explanation of their possible-world semantics
  based on the higher-order Herbrand models, TP-programs and
  PT-programs will have the same solution for their models, based on
  Interval PSAT.
\section{Probabilistic logic in time}
In this section we will investigate the generalization of a (non
temporal) probabilistic logic in time. We will consider a program
$P$ with the same set of rules, but with consecutive (in time)
fitting of the probability intervals in its p-annotated basic
formulae; the granularity of calendar can be, for example, day, weak, month, etc...\\
We will consider the following evolution of the same p-clause of a
p-program $P$, in the i-th instance of time $\textbf{t}_i$:\\
$A:[a_i,b_i] \leftarrow F_1:[a_{1,i},b_{1,i}] \wedge...\wedge
F_k:[a_{k,i},b_{k,i}]$\\
The question is: can we capture these evolutions in time of the same
p-program $P$ in a unique PT-program, by replacing the p-clauses
with the equivalent tp-clauses, where only tp-annotations are
modified. In what follows we will show that it is possible, while it
is not possible by the syntax version of TP-programs in
\cite{DeDS99T}.
\begin{definition}(\textbf{Probability Distribution Evolution})
Let $P$ be a probabilistic logic program (p-program) with a Herbrand
base $B_L$ , which in a given  instance of time $\textbf{t}$ has the
world probability distribution $PI(\textbf{t}):\textbf{2}^{B_L}
\rightarrow [0,1]$, which is a model of $P$ (satisfies all p-clauses
in $P$).
\\The complete set of these probability distributions, for all time points in
a given interval of time $\bigtriangleup_{\tau} \subseteq S_{\tau}$,
will define the mapping $PI: \bigtriangleup_{\tau} \rightarrow
([0,1]^{(\textbf{2}^{B_L})}$, which represents the \emph{evolution}
of the world probability distribution for the given p-program $P$
\emph{in this interval of time}.
\end{definition}
Let $H_F = \{r_F(\textbf{d},\textbf{t})~|~r(\textbf{d}) \in B_L$ and
$\textbf{t} \in S_{\tau}\}$ be the extended Herbrand base for all
t-atoms, derived from original atoms of the Herbrand base $B_L$ of a
p-program $P$. Let us define the subset $ D_{\bigtriangleup}$ of
Herbrand
interpretations for this new Herbrand base $H_F$ as follows:\\
 $D_{\bigtriangleup} = \{ I_F~|~ I_F:H_F \rightarrow \textbf{2}$, such that all true t-atoms in $I_F$ have \verb"the same" (distinct) time point
 $\textbf{t} \in \bigtriangleup_{\tau} \}$\\
 It is easy to verify that each $I_F \in D_{\bigtriangleup}$,
 for which the distinct time point is $\textbf{t}$, corresponds to
 some interpretation $v:B_L \rightarrow \textbf{2}$ of the p-program
 $P$ in the time point $t$. That is, holds the following bijection
 $~~~is_F: D_{\bigtriangleup}~\simeq ~\bigtriangleup_{\tau} \times \textbf{2}^{B_L}$,
 such that for any $I_F \in D_{\bigtriangleup}$, $is_F(I_F) =
 (\textbf{t},J:B_L \rightarrow \textbf{2})$,  where for any $r(\textbf{d})
 \in B_L$, $J(r(\textbf{d})) = 1~~$ if $~~I_F(r_F(\textbf{d},
 \textbf{t})) = 1$; $~~~0~$ otherwise.
Thus, the following diagram commutes\\
\begin{diagram}
   [0,1]^{(\textbf{2}^{B_L})} \times \textbf{2}^{B_L}  & \rTo^{eval}   & [0,1] \\
 \uTo^{PI} \uTo^{~~id_{\textbf{2}^{B_L}}}& \ruTo_{[PI]^{-1}} & \uTo_{[PI]^{-1}\circ is_F} \\
 \bigtriangleup_{\tau} \times ~~~\textbf{2}^{B_L}  &  \lTo^{is_F}    & D_\bigtriangleup    \\
\end{diagram}
Let us define the conservative extension $DI:\textbf{2}^{H_F}
\rightarrow [0,1]$ of the mapping $[PI]^{-1}\circ
is_F:D_\bigtriangleup \rightarrow [0,1]$, such that for any $I_F \in
D_\bigtriangleup$ it is equal to the mapping $[PI]^{-1}\circ is_F$,
and for all other $I_F \in \textbf{2}^{H_F}$ with $I_F \notin
D_\bigtriangleup$ it is equal to zero.
\begin{definition}(\textbf{Evolution PT-program}) Let $P$ be a p-program
with a Herbrand base $B_L$ with the evolution of the world
probability distribution $PI: \bigtriangleup_{\tau} \rightarrow
([0,1]^{(\textbf{2}^{B_L})}$ in the time interval $\bigtriangleup_{\tau} = [\textbf{t}_1, \textbf{t}_N]$,
 where $N$ is the cardinality of  $\bigtriangleup_{\tau}$.\\
We define the PT-program $P_{\bigtriangleup}$, denominated Evolution
PT-program in the time interval, with a Herbrand base composed by
t-atoms $H_F =
\{r_F(\textbf{d},\textbf{t})~|~r(\textbf{d}) \in B_L$, as follows:\\
each p-annotated basic formula $F:[a,b]$ of the p-program $P$, where
p-annotation in the i-th instance of time $\textbf{t}_i,~1 \leq i
\leq N$ has the value $[a_i,b_i]$, is substituted by the
correspondent tp-annotated basic formula\\ $~~~F_T:\langle
y:\textbf{t}_1\thicksim \textbf{t}_N,\{a_1,...,a_N\},
\{b_1,...,b_N\}\rangle$,\\
where $F_T$ is the formula $F$ where all atoms (with only object
variables) are replaced by equivalent t-atoms with the same object
variables and the temporal variable $y$.
\end{definition}
 Now we will demonstrate
that the whole evolution of the p-program in time can be
equivalently represented in the unique PT-program: that is, all
versions of p-programs are contained in this unique PT-program.
\begin{propo}  Let $P_{\bigtriangleup}$ be the evolution PT-program of the given p-program $P$
 with a Herbrand base $B_L$, the evolution $PI: \bigtriangleup_{\tau} \rightarrow
([0,1]^{(\textbf{2}^{B_L})}$ and its canonical extension $DI$. Then
the mapping $~KI = DI/|S_{\tau}|:\textbf{2}^{H_F}\rightarrow [0,1]$
is the model of this PT-program $P_{\bigtriangleup}$.
\end{propo}
\textbf{Proof}: As first, we will show that $KI$ is the world
probability density
function for the PT-program $P_{\bigtriangleup}$:\\
in fact, for all $I_F \in \textbf{2}^{H_F}$, $KI(I_F) =
DI(I_F)/|S_{\tau}| \geq 0$, and\\
$\sum_{I_F \in \textbf{2}^{H_F}}KI(I_F) = \sum_{I_F \in
\textbf{2}^{H_F}}DI(I_F)/|S_{\tau}| \\ =   \sum_{\textbf{t} \in
S_{\tau}} \sum_{v \in \textbf{2}^{B_L}}[PI]^{-1}(\textbf{t},v)/
|S_{\tau}|\\ = \sum_{\textbf{t} \in S_{\tau}} \sum_{v \in
\textbf{2}^{B_L}}PI(\textbf{t})(v)/|S_{\tau}| \\ = \sum_{\textbf{t}
\in S_{\tau}}1/|S_{\tau}| = 1$,\\
from the fact that $PI(\textbf{t})$ is the world probability
distribution for p-program $P$ in a time instance $\textbf{t}$ and,
consequently, satisfies the Kolmogorov axioms.\\
As consequence also $KI$ satisfies the Kolmogorov axioms for the
evolution PT-program $P_{\bigtriangleup}$, so it is its world
probability density function.\\
Now we have only to show that it is also a \emph{model}  for
$P_{\bigtriangleup}$.\\
Let us show how it works for tp-annotated basic formula
\\$(1)~~~F_T:\langle y:\textbf{t}_1\thicksim
\textbf{t}_N,\{a_1,...,a_N\}, \{b_1,...,b_N\}\rangle$.\\
 We will
reason w.r.t. each single instant of time: let  the p-annotated
basic formula $F:[a_1,b_i]$  of the p-program $P$ in the instant
time $\textbf{t}_i$ be satisfied by the world probability
distribution $PI(\textbf{t}):\textbf{2}^{B_L} \rightarrow [0,1]$.
Then we have to show that $KI$ must satisfy the formula (1) for the
instant time $\textbf{t}_i$. In fact we have that\\
$KI = (1/|S_{\tau}|)[PI]^{-1}\circ is_F:\textbf{2}^{B_L} \rightarrow
[0,1]$, thus $ KI$ is equivalent to \\$ ~(1/|S_{\tau}|) eval
\circ(PI \times id):\bigtriangleup_{\tau} \times \textbf{2}^{B_L}
\rightarrow [0,1]$, which is equivalent to disjunctive sum of
mappings\\ $~ \sum_{\textbf{t} \in
\bigtriangleup_{\tau}}PI(\textbf{t}): \textbf{2}^{B_L} \rightarrow
[0,1]$ ,\\
that is, for the instant of time $\textbf{t}_i$ it corresponds to
the mapping $PI(\textbf{t}): \textbf{2}^{B_L} \rightarrow [0,1]$
which is a model of a p-programs in the time instance $\textbf{t}_i$
and satisfies the probabilistic constraint (the probability interval
$[a_1,b_i]$, in the instance $\textbf{t}_i$) of the formula (1).\\
By structural induction we can extend the proof to all formulae and
tp-clauses in the Evolution PT-program $P_{\bigtriangleup}$.\\
%
\section{Future work and conclusion}
In this paper we defined a new syntax version for temporal
probabilistic logic programs (PT-programs) which uses explicitly the
time variable in its t-atoms, and more expressive tp-annotations for
interval probabilities, and we have shown that it can be reduced to
pure probabilistic programs, with the standard world-based
probabilistic
model theory based on Herbrand interpretations.\\
We have shown that each PT-program has the model theoretic semantics
equivalent to the model theoretic semantics of TP-programs defined
in \cite{DeDS99T}, and explain the reasons why the fixed point
semantics in \cite{DeDS99T} is generally non valid. Moreover, we
explain also the meaning of "threads', use for possible worlds in
\cite{DeDS99T}, in terms of higher-order Herbrand models of
PT-programs.\\
In the significant example for a kind of version-system for ordinary
p-programs, we have shown that the probabilistic program evolution
in time, by modifying p-annotations for its basic formulae, can be
embedded into the unique PT-program with the same set of clauses, by
defining tp-annotations of its basic formulae in order to support
this probability-interval modifications of the original p-program.
Such feature can not be supported by the original syntax for
TP-programs presented in \cite{DeDS99T}.
 We used the reduction of PT-programs into ordinary p-programs to define the
complexity for the consistency (NP-complete) and entailment problem
(co-NP-complete) in the general case of PT-programs (it must hold
also for PT-programs defined in \cite{DeDS99T}, because they have
the same model theoretic semantics).\\
By incorporation of the time variable into all atoms of PT-programs,
we obtain that the facts of PT-programs define the TP-tuples of
Temporal Probabilistic Databases, so that the whole PT-program can
be seen as a kind of virtual TP-database, and a kind of
Probabilistic-DATALOG. The relationship with
many-valuedness and intensionality is presented in \cite{Majk09P}.\\
   The future work  will be dedicated to explore such PT-programs for TP-databases,
   especially  a query-answering in such virtual TP-databases. The
   query-answering in such TP-databases (i.e., PT-programs) is
   closely related to the entailment in PT-programs: a ground query
   $F:\mu$   is just the entailment of this formula from a given
   PT-program; a query with variables will return with a set of
   ground formulae, each one entailed from the PT-program.\\
   Thus, the high complexity of query answering (co-NP-complete) can
   be a problem for the big TP-databases. If we consider a
   PT-program with only one binary atom and 10 constants, the number
   of possible worlds (that is, the number of variables for Interval SAT) is equal to $2^{100} \approx 10^{30}$
   !\\
   Thus we need more investigation, in order to reduce this
   complexity. One of them is to reduce the complexity of
   PT-programs, for example, the \emph{simple strict} programs (see for more details in
   \cite{DeDe05})
   have the polynomial complexity, but are to much strong simplification
   for real problems: we need to investigate some minor syntax
   restriction for PT-programs but with in-practice acceptable
   query-answering complexity.\\
   The other possibility is to consider only a strict subset of models of
   PT-programs for the \emph{plausible query-answering}, as usually
   applied in query-answering in databases with subset of preferred
   inconsistency repairings \cite{Majk05p}, or in non monotonic logic
   programming. The extremal point of reductions, to the unique
   preferred model, can be applied if we chose, for example, to use
   the model with \emph{maximum entropy} (MA) already stated in the work by Nilsson \cite{Nils86}. The maximum entropy model
   is the unique probabilistic interpretation $KI:\textbf{2}^{H_F} \rightarrow
   [0,1]$, which is a model of a PT-program $P$ with a Herbrand base $H_F$, and has the greatest
   entropy among all the models of $P$, where the entropy of $Pr$,
   denoted $H(Pr)$, is defined by:\\
   $~~~~~~~~~~H(KI) = - \sum_{I_F \in \textbf{2}^{H_F}} KI(I_F)\cdot log
   KI(I_F)$.\\
   Principle of maximum entropy may be taken   to compute
   degrees of belief of formulae \cite{GrHK94}, and it is shown in \cite{PaVe90} for the
   consistent probabilistic inference. This method applied to
   probabilistic logic programming \cite{LuKe99}, based on \emph{conditional}
   probabilistic clauses, has shown that reduces the original
   entropy maximizations to relatively small optimization problems,
   which can easily be solved by existing MA-technology.
   %
%


\bibliographystyle{IEEEbib}
\bibliography{mydb}


%
\end{document}